\documentclass[12pt,a4paper]{article}
\pdfoutput=1
\usepackage{jheppub}
\usepackage{amsmath,amssymb,bbm,slashed,cancel}
\usepackage{graphicx}
\usepackage{slashed}
\usepackage{hyperref}
\usepackage{subcaption}
\usepackage{dsfont}
\usepackage{stackengine}
\usepackage[utf8]{inputenc}
\usepackage{multirow}
\usepackage{physics}
\usepackage{mathtools}
\usepackage[normalem]{ulem}

\usepackage{comment}

\usepackage{colortbl,colordvi,xcolor,color}

\usepackage{array,multirow}
\usepackage{booktabs}
\usepackage{float}

\allowdisplaybreaks
\usepackage{cprotect}

\newcommand{\eg}{{\em e.g.}}
\newcommand{\ie}{{\em i.e.}}
\newcommand\unit[1]{\,\mathrm{#1}}

\newcommand\GeV{\unit{GeV}}
\newcommand\TeV{\unit{TeV}}
\newcommand\pt{p_{\rm T}}

\makeatletter
\gdef\@fpheader{}
\makeatother

\preprint{KEK--TH--2595}

\title{Heavy photophobic ALP at the LHC}
\author[a]{Masashi Aiko}
\author[a,b]{Motoi Endo}
\author[a,c]{and Kåre Fridell}

\affiliation[a]{KEK Theory Center, Tsukuba, Ibaraki 305--0801, Japan}
\affiliation[b]{Graduate Institute for Advanced Studies, SOKENDAI, Tsukuba, Ibaraki 305--0801, Japan}
\affiliation[c]{Department of Physics, Florida State University, Tallahassee, FL 32306, USA}

\emailAdd{maiko@post.kek.jp}
\emailAdd{motoi.endo@kek.jp}
\emailAdd{kare.fridell@kek.jp}

\abstract{
We study the photophobic ALP model in high-mass regions under LHC Run-II.
Since the ALP is predominantly coupled with electroweak gauge bosons such as $ZZ$, $WW$, and $Z\gamma$, and less with di-photon, the model may be probed via multi-boson final-state processes.
We find that on-shell ALP productions with $Z\gamma$ final states currently provide the best sensitivities for $m_{a} > 40\GeV$.
}

\begin{document}
\maketitle
\flushbottom

\renewcommand{\thefootnote}{\#\arabic{footnote}}
\setcounter{footnote}{0}

\section{Introduction} \label{sec: intro}

Pseudo-scalar bosons are one of the most attractive candidates for new particles.
They often appear in models with extended scalar sectors. 
In particular, an axion-like particle (ALP) has been widely studied. 
Its mass is assumed to be generated by explicit violations of global symmetries, and couplings with gauge bosons in the Standard Model (SM) are characterized by remnant shift symmetries.

We consider photophobic ALPs, whose masses are large, and coupling to di-photon is suppressed (cf., Ref.~\cite{Craig:2018kne}).
Such a model has attractive features\footnote{
Although the photophobia is not realized generically, it may arise in different scenarios (cf., Refs.~\cite{Hook:2016mqo, Craig:2018kne}).
An explicit model is given in Ref.~\cite{Craig:2018kne}.
}, such as relaxing experimental constraints (see Refs.~\cite{Mimasu:2014nea, Jaeckel:2015jla, Jaeckel:2012yz, Bauer:2017ris, Bauer:2018uxu, Brivio:2017ije, Florez:2021zoo, Wang:2021uyb, dEnterria:2021ljz, Knapen:2016moh, CMS:2018erd, ATLAS:2020hii, Bonilla:2022pxu, Biswas:2023ksj,Liu:2023bby,Liu:2021lan}). 
Nonetheless, the low-mass regions of its parameter space are restricted by cosmology, flavor, and LEP experiments~\cite{Izaguirre:2016dfi, Alonso-Alvarez:2018irt, Gavela:2019wzg, Guerrera:2021yss, Bauer:2021mvw, Guerrera:2022ykl, Craig:2018kne, Ebadi:2019gij}, while the high-mass regions have been less constrained~\cite{Craig:2018kne, Bonilla:2022pxu, Aiko:2023trb, Aiko:2023nox}. 
Furthermore, the global fit to the electroweak precision data may be improved~\cite{Aiko:2023trb}.

In this paper, we study heavy photophobic ALPs in light of the LHC Run-II results.
In the previous work~\cite{Craig:2018kne}, on-shell productions of ALPs have been investigated by using LHC data at the center-of-mass energy $\sqrt{s}=8\TeV$ with an integrated luminosity $\int\!\mathcal{L}\,dt\simeq 20\,{\rm fb}^{-1}$.
The sensitivities are degraded in the high-mass regions because the production cross section and/or luminosity was not large enough. 
In another work~\cite{Bonilla:2022pxu}, non-resonant signals of off-shell ALPs were explored for ALP masses up to $100\GeV$. 
Above this mass range, on-shell productions become relevant. 
Hence, we revisit LHC searches for on-shell ALP productions especially in high-mass regions by employing the latest LHC Run-II data.
Since the ALP is coupled mainly with the electroweak gauge bosons ($ZZ$, $WW$, and $Z\gamma$), we investigate multi-boson final-state processes, $pp\to WWW$, $pp\to Z\gamma\to \nu\bar{\nu}\gamma$, and $pp\to Z\gamma\to \ell^+\ell^-\gamma$. 

This paper is organized as follows.
In Sec.~\ref{sec: ALP_model}, we briefly introduce the photophobic ALP model. 
In Sec.~\ref{sec: MC_Analysis}, we explain how to reinterpret the LHC data for constraints on the ALP coupling, and show numerical results in Sec.~\ref{sec: numerical results}.
We conclude in Sec.~\ref{sec: conclusion}.
Lastly, although the original photophobic ALP model does not include a direct ALP-gluon coupling, we analyze the case when the coupling is introduced in Appendix~\ref{sec: ALP-gluon coupling}.

\section{Model} \label{sec: ALP_model}

The photophobic ALP is assumed to be coupled to the ${\rm SU(2)}_L$ and ${\rm U(1)}_Y$ gauge bosons.
The other direct couplings with SM particles are suppressed.
The ALP Lagrangian is shown as~\cite{Georgi:1986df}
\begin{align}
\mathcal{L}_{\rm ALP} &=
\frac{1}{2}\partial_{\mu}a\partial^{\mu}a
-\frac{1}{2} m_{a}^{2} a^{2}
-c_{WW}\frac{a}{f_{a}}W_{\mu\nu}^{a}\widetilde{W}^{a\mu\nu}
-c_{BB}\frac{a}{f_{a}}B_{\mu\nu}\widetilde{B}^{\mu\nu},
\label{eq: Lagrangian}
\end{align}
where $f_{a}$ is the ALP decay constant, and where we assume that $m_a \lesssim f_a$ is satisfied, see Refs.~\cite{Aiko:2023trb, Aiko:2023nox} for our conventions.
After electroweak symmetry breaking, the interactions are generally expressed as
\begin{align}
\mathcal{L}_{\mathrm{int}}
&=
-\frac{1}{4}g_{a\gamma\gamma}aF_{\mu\nu}\widetilde{F}^{\mu\nu}
-\frac{1}{2}g_{a\gamma Z}aZ_{\mu\nu}\widetilde{F}^{\mu\nu} \notag \\
&\quad
-\frac{1}{4}g_{aZZ}aZ_{\mu\nu}\widetilde{Z}^{\mu\nu}
-\frac{1}{2}g_{aWW}aW_{\mu\nu}^{+}\widetilde{W}^{-\mu\nu}
+ \cdots.
\label{eq: effective_coupling}
\end{align}
The coupling constants are related to $c_{WW}$ and $c_{BB}$ as
\begin{align}
g_{a\gamma\gamma} &= \frac{4}{f_{a}}\qty(s_{W}^{2}{c}_{WW}+c_{W}^{2}{c}_{BB}), 
\label{eq: gaAA} \\
g_{aZ\gamma} &= \frac{2}{f_{a}}\qty(c_{WW}-c_{BB})s_{2W}, 
\label{eq: gaZA} \\
g_{aZZ} &= \frac{4}{f_{a}}\qty(c_{W}^{2}c_{WW}+s_{W}^{2}c_{BB}), 
\label{eq: gaZZ} \\
g_{aWW} &= \frac{4}{f_{a}}c_{WW},
\label{eq: gaWW}
\end{align}
where $c_{W} = \cos{\theta_{W}}$, $s_{W} = \sin{\theta_{W}}$, $c_{2W} = \cos{2\theta_{W}}$, and $s_{2W} = \sin{2\theta_{W}}$ with the Weinberg angle $\theta_{W}$.
In the photophobic model, the ALP coupling to di-photon is absent at the tree level, \ie, $s_{W}^{2}{c}_{WW}+c_{W}^{2}{c}_{BB} = 0$ is satisfied, yielding $g_{a\gamma\gamma}=0$.
Then, $g_{aZ\gamma}$, $g_{aZZ}$ are denoted in terms of $g_{aWW}$ as
\begin{align}
g_{aZ\gamma} = \frac{s_{W}}{c_{W}}g_{aWW}\qc \label{eq: gaZgamma}
g_{aZZ} = \frac{c_{2W}}{c_{W}^{2}}g_{aWW}.
\end{align}

In the model, there are no direct ALP couplings with SM particles other than $ZZ$, $WW$, and $Z\gamma$.
In particular, no direct ALP-gluon coupling $g_{aGG}$ or ALP-fermion couplings $g_{aff}$ are introduced.
In UV models (see \eg, Ref.~\cite{Craig:2018kne}), $g_{a\gamma\gamma}=g_{aGG}=g_{aff}=0$ is realized at a 
high scale.
If the ALP is massless, as a result of the shift symmetry, $g_{a\gamma\gamma}$ and $g_{aGG}$ are not generated via renormalization group evolution, but $g_{aff}$ is generated at the one-loop level (see  Ref.~\cite{Bauer:2020jbp,Chala:2020wvs}).
On the other hand, non-zero ALP mass generates $g_{a\gamma\gamma}$ and $g_{aGG}$ at the one- and two-loop levels, respectively~\cite{Bauer:2017ris}, since the symmetry is softly broken.
Nonetheless, such loop-induced $g_{a\gamma\gamma}$ and $g_{aGG}$ are considered to be less effective for LHC searches.\footnote{
Although the original setup of the photophobic ALP model predicts $g_{aGG} = 0$, its presence may affect the collider signatures significantly. 
We investigate LHC signatures when $g_{aGG}$ is turned on at tree level in Appendix \ref{sec: ALP-gluon coupling}.
} 

\subsection{Decay of ALP} \label{subsec: decay}

The ALP decays into a pair of SM gauge bosons.
The partial decay widths for $a\to V_{i}V_{j}\ (V_{i,j}=\gamma,Z,W^{\pm})$ are obtained as~\cite{Bauer:2017ris, Craig:2018kne, Bonilla:2021ufe}
\begin{align}
    \Gamma(a\to V_{i}V_{j}) = \frac{m_{a}^{3}}{32\pi(1+\delta_{ij})}\lambda^{3/2}
    \qty(\frac{m_{V_{i}}^{2}}{m_{a}^{2}}, \frac{m_{V_{j}}^{2}}{m_{a}^{2}})\abs{g_{aV_{i}V_{j}}^{\rm eff}}^{2},
    \label{eq: decay_a_VV}
\end{align}
with $\lambda(x,y) = (1-x-y)^{2}-4xy$.
Here $m_{V}$ is the gauge-boson mass ($m_\gamma=0$).
Note that $\delta_{ij}=0$ for $a\to Z\gamma$ and $a\to W^+ W^-$.

The ALP coupling to di-photon $g^{\rm eff}_{a\gamma\gamma}$ is induced by loop corrections with $g_{aWW}$ in Eq.~\eqref{eq: gaWW} as~\cite{Bauer:2017ris}
\begin{align}
    g_{a\gamma\gamma}^{\rm eff} = \frac{2\alpha}{\pi} g_{aWW}B_{2}(\tau_{W}),    
\end{align}
where $\alpha\equiv e^{2}/(4\pi)$ with the QED coupling $e=g s_{W}$.
The loop function $B_{2}$ is defined as
\begin{align}
    B_{2}(\tau) = 1-(\tau-1)f^{2}(\tau),
\end{align}
where $f(\tau)$ is given by
\begin{align}
    f(\tau) =
    \left\{\mqty{
        \arcsin{\frac{1}{\sqrt{\tau}}}, \\
        \frac{\pi}{2}+\frac{i}{2}\log{\frac{1+\sqrt{1-\tau}}{1-\sqrt{1-\tau}}},
    }\right.\qquad
    \mqty{
        \text{for}~\tau\geq 1\\
        \text{for}~\tau< 1
    },
\end{align}
with $\tau_{W}=4m_{W}^{2}/m_{a}^{2}$.
On the other hand, it is sufficient for us to evaluate the decay widths for $a\to Z\gamma,\, ZZ,\, W^{+}W^{-}$ at the tree level, \ie, $g_{aV_{i}V_{j}}^{\rm eff} = g_{aV_{i}V_{j}}$.

When the ALP is lighter than the $Z$ boson,
the ALP decays into either a pair of SM fermions, $a \to f\bar f$, or the fermions with a photon, $a \to Z^{*}\gamma\to f\bar f \gamma$, by exchanging an off-shell $Z$ boson. 
Since the ALP does not couple with SM fermions directly, the former proceeds via radiative corrections.
On the other hand, even though the latter is a three-body decay, it proceeds at the tree level, and thus, its decay width can be comparable to the former. 

For $m_a \ll m_{Z}$, the partial decay width of $a \to Z^{*}\gamma\to f\bar f \gamma$ is obtained as
\begin{align}
 \Gamma(a \to Z^{*}\gamma\to f\bar f \gamma) =
 N_{c}^{f}\frac{g_{Z}^{2}g_{aZ\gamma}^{2}}{30720\pi^{3}}
 \qty[(g_{V,f})^{2}+(g_{A,f})^{2}]\frac{m_{a}^{7}}{m_{Z}^{4}},
\end{align}
where $N_{c}^{f} = 1~ (3)$ is the color factor for leptons (quarks), and $g_Z = g/c_{W}$ is the $Z$-boson coupling constant.
The vector and axial form-factors of $Z$ boson are defined as $g_{V,f} = I_{3}^f - 2\, Q_f s_W^2$ and $g_{A,f} = I_{3}^f$.
The formula valid for any $m_a$ is provided in Ref.~\cite{Aiko:2023trb}.

The ALP decays into a pair of leptons and heavy quarks, $a\to f\bar{f}$, via gauge-boson loops.
The partial decay widths are obtained as~\cite{Bauer:2017ris, Bauer:2020jbp}
\begin{align}
    \Gamma(a\to f\bar{f}) =
    N_{c}^{f}\frac{m_{a}m_{f}^{2}}{8\pi}
    \abs{g_{aff}^{\rm eff}}^{2}
    \sqrt{1-\frac{4m_{f}^{2}}{m_{a}^{2}}}.
\end{align}
The effective coupling induced by radiative corrections is given by
\begin{align}
    g_{aff}^{\rm eff} &=
    \frac{3}{4s_{W}^{2}}\frac{\alpha}{4\pi}g_{aWW} \ln{\frac{\Lambda^{2}}{m_{W}^{2}}}
    \notag \\ &\quad
    +\frac{3}{s_{W}c_{W}}\frac{\alpha}{4\pi}g_{aZ\gamma}Q_{f}
        \qty(I_{3}^{f}-2Q_{f}s_{W}^{2})
        \ln{\frac{\Lambda^{2}}{m_{Z}^{2}}}
    \notag \\ &\quad
    +\frac{3}{s_{W}^{2}c_{W}^{2}}\frac{\alpha}{4\pi}g_{aZZ}
        \qty(Q_{f}^{2}s_{W}^{4}-I_{3}^{f}Q_{f}s_{W}^{2}+\frac{1}{8})
        \ln{\frac{\Lambda^{2}}{m_{Z}^{2}}},
    \label{eq: cff_eff}
\end{align}
where $Q_{f}$ and $I_{3}^{f}$ are the electric charge and the weak isospin of the fermion $f$, and $\tau_{f}=4m_{f}^{2}/m_{a}^{2}$.
Here, we keep the terms enhanced by $\ln{(\Lambda^2/m_f^2)}$ or $\ln{(\Lambda^2/m_{V}^2)}$ $(V=Z, W)$.
We take the cutoff scale $\Lambda$ to be $4\pi f_{a}$ in this paper.
Subleading terms are given in Refs.~\cite{Bauer:2017ris, Bauer:2020jbp, Bonilla:2021ufe}, though they are irrelevant here (cf., Ref.~\cite{Arias-Aragon:2022iwl}).

On the other hand, the inclusive rate for the ALP decaying into light hadrons is shown as (cf., Ref.~\cite{Bauer:2017ris})
\begin{align}
    \Gamma(a\to{\rm hadrons}) &=
    32\pi\alpha_{s}^{2}m_{a}^{3}\qty(1+\frac{83}{4}\frac{\alpha_{s}}{\pi})\abs{g_{aGG}^{\rm eff}}^{2}
    +\sum_{f=u,d,s}\Gamma(a\to f\bar{f}),
    \label{eq: aToHadron}
\end{align}
where $m_{a}\gg\Lambda_{\rm QCD}$ is assumed. 
The effective coupling with gluons is given by
\begin{align}
    g_{aGG}^{\rm eff}=\frac{1}{32\pi^{2}}\sum_{f=u,d,s}g_{aff}^{\rm eff}. \label{eq: gagg-eff}
\end{align}
Note that the above expression is not valid for $m_{a} \lesssim 3~{\rm GeV}$, where the perturbative QCD is not applicable, and various hadronic decay channels appear (cf., Refs.~\cite{Bauer:2017ris, Aloni:2018vki}).

\begin{figure}[t]
 \centering
 \includegraphics[scale=0.6]{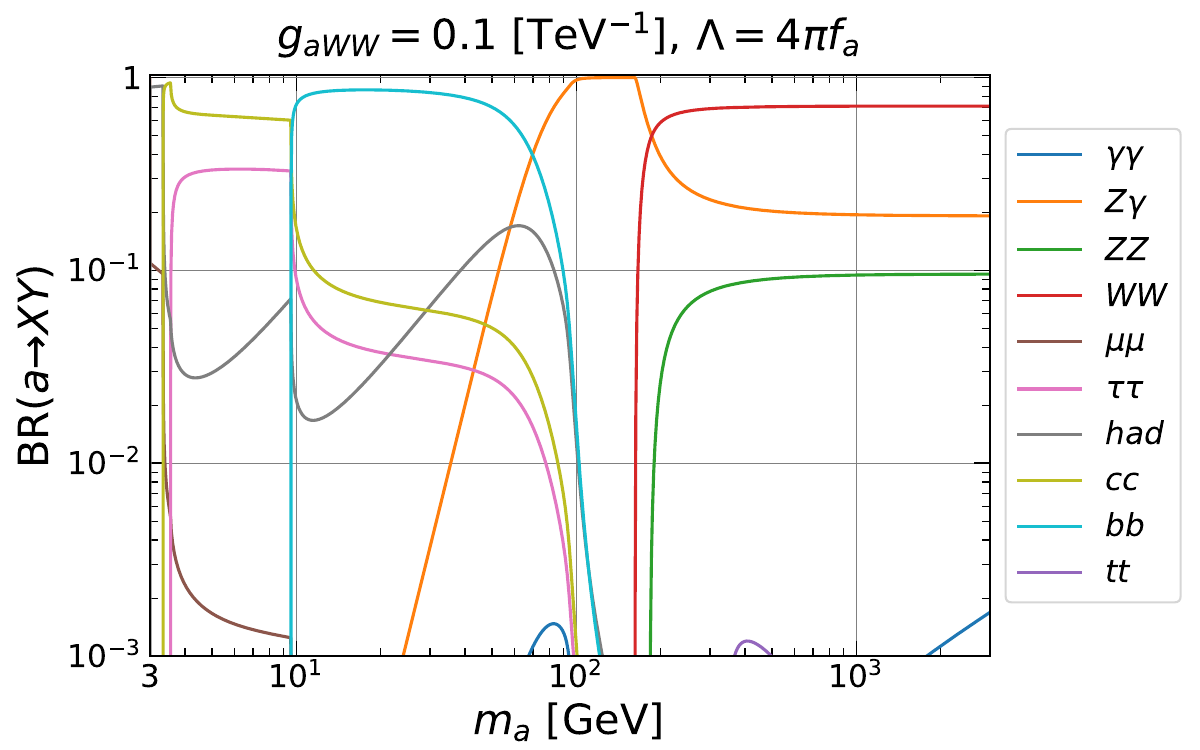}
 \caption{Branching ratios of the photophobic ALP satisfying $g_{a\gamma\gamma}=g_{aGG}=g_{aff}=0$ at tree level. 
 Here, we take $g_{aWW} = 0.1\TeV^{-1}$ and $\Lambda = 4\pi f_{a} = 4\pi\times10\TeV$.}
 \label{fig: ALP_decay_gaAA0gaWW01gaGG0}
\end{figure}

We show the branching ratios of the photophobic ALP decay in Fig.~\ref{fig: ALP_decay_gaAA0gaWW01gaGG0}.
We set $g_{aWW}=0.1\TeV^{-1}$ and the cutoff scale $\Lambda = 4\pi f_{a} = 4\pi\times10\TeV$.
For $m_{a}\lesssim m_{Z}$, the decay is dominated by loop-induced decays into a pair of fermions, and by the tree-level decays into a pair of gauge bosons for $m_{a}>m_{Z}$.
In particular, $a \to WW$ and/or $Z\gamma$ becomes dominant for $m_{a} \gtrsim 100\GeV$.
Note that the neutrino channels are extremely suppressed. 
Also, $a\to Zh$ and $a\to hh$ vanish in our setup.

\section{LHC signatures} \label{sec: MC_Analysis}

Light ALPs are constrained by cosmology and precision/collider experiments. 
If the ALP mass is smaller than $\sim1\GeV$, the model is severely limited by cosmological measurements~\cite{Craig:2018kne}, while heavier ALPs are constrained by flavor and collider experiments.
In particular, heavy meson decays are sensitive to $g_{aWW}$ for $m_{a}\lesssim 4.8\GeV$.
The current status is summarized in Ref.~\cite{Aiko:2023trb}.
For $4.8\GeV\lesssim m_{a} \lesssim m_{Z}$, LEP results have provided the leading constraints via searches for $Z$-boson invisible decays and $Z\to \gamma a\to \gamma jj$.
The detailed analysis can be found in Ref.~\cite{Craig:2018kne}.

In this paper, we focus on heavy ALPs.
For $m_{a} \gtrsim m_{Z}$ the LHC measurements are sensitive to ALP contributions and have provided constraints on $g_{aWW}$.
In the previous work~\cite{Craig:2018kne}, on-shell ALP productions have been studied extensively with the LHC results at $\sqrt{s}=8\TeV$ and $\int\!\mathcal{L}\,dt\simeq 20\,{\rm fb}^{-1}$.
The sensitivities are degraded rapidly as the ALP mass is increased because the production cross section/luminosity is not large enough. 
In the following, we reinterpret the latest LHC results obtained at $\sqrt{s}=13\TeV$.
In particular, we analyze $pp\to WWW$, $pp\to Z\gamma\to \nu\bar{\nu}\gamma$, and $pp\to Z\gamma\to \ell^+\ell^-\gamma$, via on-shell ALP production.\footnote{
There are briefly two approaches to studying ALP contributions at the LHC: searches for new resonances in event distributions and for event number excesses after kinematical cuts. 
For the former, among the ALP decay channels, $a \to Z\gamma$ provides a sensitive probe and will be studied in this paper. See Refs.~\cite{CMS:2021xor,CMS:2021itu} for $a \to ZZ, WW$.
For the latter approach, we analyze $pp\to WWW$ and $pp\to Z\gamma\to \nu\bar{\nu}\gamma$ proceeding via $a \to WW$ and $Z\gamma$ respectively. 
Although other processes such as $pp \to WWZ, WZZ$, and $ZZZ$ (see Ref.~\cite{CMS:2020hjs}) may contribute, their limits on the ALP models are expected to be weaker than those obtained in this paper.
}

To generate the event samples, we employ \verb|MadGraph5_aMC@NLO 3.5.2|~\cite{Alwall:2014hca} interfaced with \verb|PYTHIA 8.3|~\cite{Bierlich:2022pfr} for parton showering and hadronization.
The ALP model is implemented by using a UFO model file generated with \verb|FeynRules 2.3|~\cite{Alloul:2013bka}.
Detector simulations are performed by \verb|DELPHES 3.5.0|~\cite{deFavereau:2013fsa}.

\subsection{Statistical method}

The signals are investigated by comparing the number of ALP signal events ($n^{\rm sig}$) with the number of observed events ($n^{\rm obs}$) and the expected number of background events ($n^{\rm bkg}$). 
For $pp\to WWW$ and $pp\to Z\gamma\to \nu\bar{\nu}\gamma$, we evaluate the upper limits by using a modified frequentist approach, which is also called the ${\rm CL}_{s}$ method~\cite{Junk:1999kv, Read:2002hq}.\footnote{
Such a prescription is not necessary for resonance searches via $pp\to Z\gamma\to \ell^+\ell^-\gamma$. See Sec.~\ref{sec:Zllgamma} for details.
}
The events are assumed to follow Poisson distributions, and the $p$-value and ${\rm CL}_{s}$ are defined as
\begin{align}
p_{\alpha} = \sum_{n=0}^{n^{\rm obs}} \frac{\alpha^{n}e^{-\alpha}}{n!}\qc
{\rm CL}_{s} = \frac{p_{s+n^{\rm bkg}}}{p_{n^{\rm bkg}}}.
\label{eq:CLs}
\end{align}
The upper limit on the number of signal events at the 95\% level, $n^{\rm sig}_{\rm up}$, is determined by solving ${\rm CL}_{s} = 0.05$.
Then, the upper bound on $g_{aWW}$ is obtained from
\begin{align}
n^{\rm sig} = \sigma
\cdot{\cal L}_I\cdot A \times \epsilon
< n^{\rm sig}_{\rm up},
\label{eq:nsigup}
\end{align}
where ${\cal L}_I$ is an integrated luminosity, $A$ is the acceptance, and $\epsilon$ accounts for the efficiency.
The production cross section $\sigma$ is a function of $m_a$ and $g_{aWW}$. 
For a given ALP mass, $\sigma$ scales as $g_{aWW}^2$ when the process is dominated by on-shell ALP productions. 
In practice, signal categories are defined by imposing cuts on the events, and an upper limit is obtained in each signal category.
Among the categories, the most stringent constraint is adopted as the result.

There are two comments in order.
First, although the ${\rm CL}_{s}$ method does not always provide a real confidence interval, it is useful especially to accommodate the case where an experimental result appears to contain fewer events than the expected number of background events due to a fluctuation, which is likely to lead to too strong exclusions. 
Such a situation occurs in some of the signal categories for $pp\to WWW$ and $pp\to Z\gamma\to \nu\bar{\nu}\gamma$. 
Next, Eq.~\eqref{eq:CLs} does not take account of systematic uncertainties. 
In the literature, they are not provided explicitly for each signal category. 
Besides, although there are potential correlations among the categories, such information is not shown, and we therefore omit them for simplicity. 
As explained below, the validity of the above method is checked by deriving limits on new physics models that are explored in Refs.~\cite{CMS:2019mpq, ATLAS:2021pdg} and comparing the results with those in the literature.

\subsection{CMS \texorpdfstring{$W^{\pm}W^{\pm}W^{\mp}$}{} search} \label{sec:WWW}

\begin{figure}[t]
 \centering
 \includegraphics[scale=0.25]{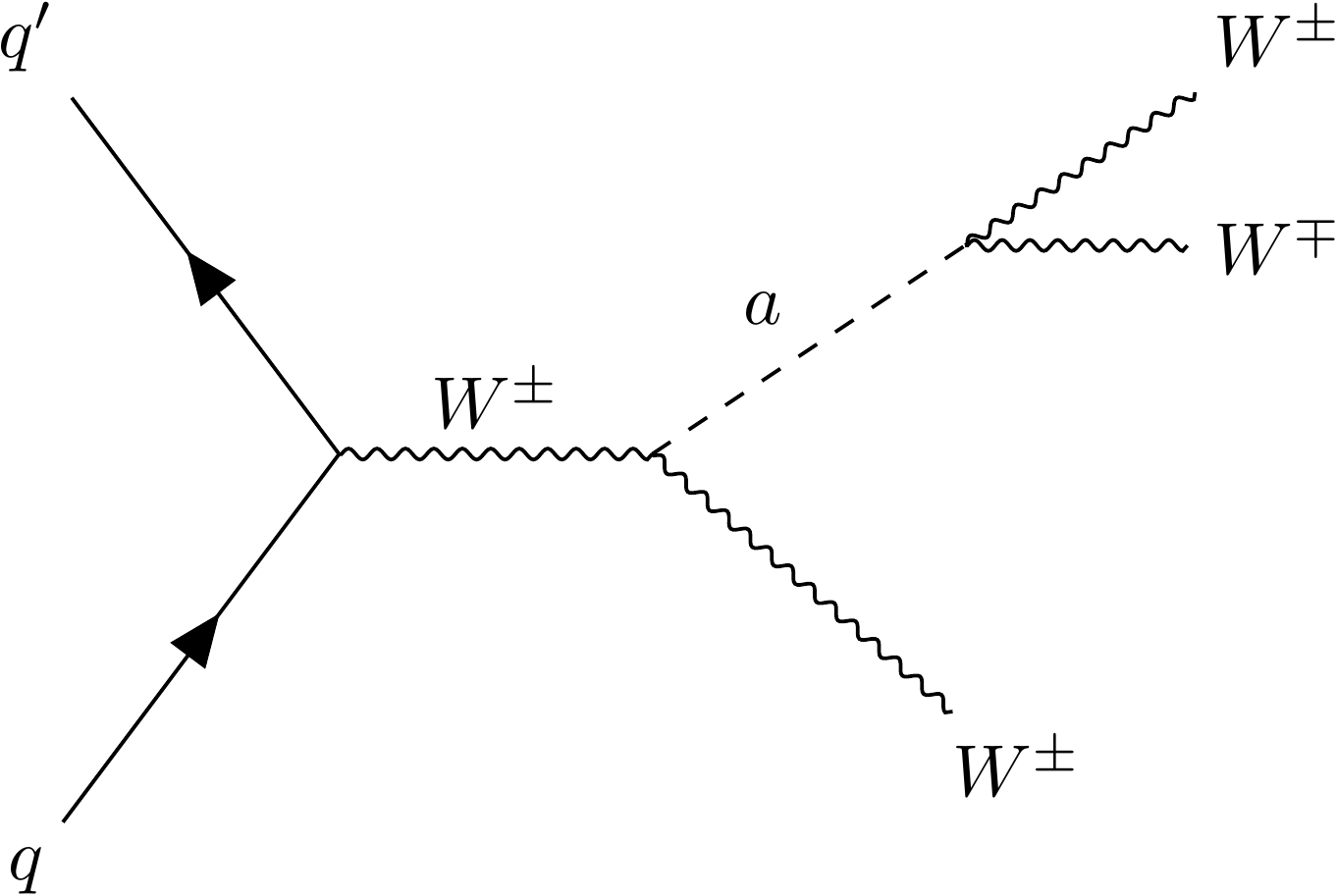} 
 \caption{Feynman diagrams for the $pp\to WWW$ production.}
 \label{fig: pp_www}
\end{figure}

The analysis is based on a data sample from the CMS collaboration at $\sqrt{s}=13\TeV$ with ${\cal L}_I = 35.9\,{\rm fb}^{-1}$~\cite{CMS:2019mpq}.\footnote{ \label{footnote: pp_www}
In this paper, we do not use the results based on the LHC Run-II data samples with ${\cal L}_I = 137\,{\rm fb}^{-1}$ (CMS)~\cite{CMS:2020hjs} and $139\,{\rm fb}^{-1}$ (ATLAS)~\cite{ATLAS:2022xnu}, because they utilize multivariate techniques and it is not straightforward to implement them into our analysis. 
}
The signal process is $pp\to W^{\pm}a \to W^{\pm}W^{\pm}W^{\mp}$ (Fig.~\ref{fig: pp_www}).
In the event generation, we do not restrict the ALP and $W$ boson to being produced on-shell, and their off-shell contributions are also taken into account.
Owing to decay channels of the $W$ bosons, two classes of signal events are considered in the experimental analysis: final states with two same-sign (SS) leptons and those with at least two jets and three leptons (3$\ell$).

\begin{table}[t]
\scriptsize
\centering
\begin{tabular}{l c c c}
\hline
 & $e^{\pm}e^{\pm}$ & $e^{\pm}\mu^{\pm}$ & $\mu^{\pm}\mu^{\pm}$ \\
\hline
Signal leptons & \multicolumn{3}{c}{2 same-sign leptons with $\pt>25\GeV$} \\
Additional leptons &  \multicolumn{3}{c}{No additional lepton} \\
Jets &  \multicolumn{3}{c}{At least two jets with $\pt>30\GeV$ and $\abs{\eta}<2.5$} \\
b-tagged jets &  \multicolumn{3}{c}{No b-tagged jet} \\
tau-tagged jets & \multicolumn{3}{c}{No tau-tagged jets} \\
\multirow{2}{*}{$m_{jj}$ (dijet mass of jets closest in $\Delta R$)} & \multicolumn{3}{c}{$65 < m_{jj} < 95\GeV$ ($m_{jj}$-in)} \\
 &  \multicolumn{3}{c}{$\abs{m_{jj}-80\GeV}\geq15\GeV$ ($m_{jj}$-out)} \\
$m_{JJ}$ (dijet mass of leading jets) & \multicolumn{3}{c}{$<400\GeV$} \\
$\Delta\eta$ of two leading jets & \multicolumn{3}{c}{$<$1.5} \\
$\pt^{\rm miss}$ &  $>60\GeV$ & $>60\GeV$ & $>60\GeV$ ($m_{jj}$-out) \\
$m_{\ell\ell}$ & $>40\GeV$ & $>30\GeV$ & $>40\GeV$ \\
$m_{\ell\ell}$  & $\abs{m_{\ell\ell}-m_{Z}} > 10\GeV$ & --- & --- \\
$m_{T}^{\rm max}$  & --- & $>90\GeV$ & --- \\
\hline
\end{tabular}
\caption{Kinematical cuts for the SS final states based on Ref.~\cite{CMS:2019mpq}, which contains two same-sign leptons and at least two jets.}
\label{tab:SS}
\end{table}

\begin{table}[t]
\scriptsize
\centering
\begin{tabular}{l c c c}
\hline
 & 0\,SFOS & 1\,SFOS & 2\,SFOS \\
\hline
\multirow{2}{*}{Signal leptons} & \multicolumn{3}{c}{3 tight leptons with $\pt>25/20/20\GeV$} \\
                                & \multicolumn{3}{c}{and charge sum = $\pm1$e} \\
Additional leptons &  \multicolumn{3}{c}{No additional rejection lepton} \\
Jets &  \multicolumn{3}{c}{At most one jet with $\pt>30\GeV$, $\abs{\eta}<5$} \\
b-tagged jets & \multicolumn{3}{c}{No b-tagged jets} \\
tau-tagged jets & \multicolumn{3}{c}{No tau-tagged jets} \\
$\pt(\ell\ell\ell)$ & --- & $>60\GeV$ & $>60\GeV$ \\
$\Delta\phi(\pt(\ell\ell\ell), \pt^{\rm miss})$ & \multicolumn{3}{c}{$>2.5$} \\
$\pt^{\rm miss}$ & $>30\GeV$ & $>45\GeV$ & $>55\GeV$ \\
$m_{T}^{\rm max}$ &  $>90\GeV$ & --- & --- \\
$m_{T}^{\rm 3d}$ & --- & $>90\GeV$ & --- \\
SF lepton mass & $>20\GeV$ & --- & --- \\
Dielectron mass  & $\abs{m_{ee}-m_{Z}} > 15\GeV$ & --- & --- \\
\multirow{2}{*}{$m_{\rm SFOS}$} & \multirow{2}{*}{---}  & $\abs{m_{\rm SFOS}-m_{Z}}>20\GeV$ & $\abs{m_{\rm SFOS}-m_{Z}}>20\GeV$ \\
 &  & and $m_{\rm SFOS}>20\GeV$ & and $m_{\rm SFOS}>20\GeV$ \\
$m_{\ell\ell\ell}$ &  \multicolumn{3}{c}{$\abs{m_{\ell\ell\ell}-m_{Z}}>10\GeV$} \\
\hline
\end{tabular}
\caption{Kinematical cuts for the $3\ell$ final events based on Ref.~\cite{CMS:2019mpq}, which contains exactly three leptons.}
\label{tab:3l}
\end{table}

The event reconstruction and event selection criteria are based on Ref.~\cite{CMS:2019mpq}.
For the detector simulations, we modify the default \verb|DELPHES| card for CMS as follows.
Electron and muon candidates which satisfy $p_{T}>10\GeV$ and $\abs{\eta}<2.4$ are selected.
We discard the electron candidates with $1.4<\abs{\eta}<1.6$ and assume 95\% and 85\% efficiencies for those with $\abs{\eta}<1.4$ and $1.6 < \abs{\eta} < 2.4$, respectively.
Jet clustering is performed in the anti-$k_{T}$ algorithm~\cite{Cacciari:2008gp} with distance parameter $R=0.4$, and jets with $p_{T}>20\GeV$ and $\abs{\eta}<5$ within $\Delta R<0.4$ are selected.
The efficiency to select $b$-quark jets is set to 80\%. 
For other variables, we use the default values.

The kinematical cuts are summarized in Tables~\ref{tab:SS} and \ref{tab:3l} for SS and $3\ell$ events, respectively.
The SS events are classified into six signal categories by the invariant mass of two jets $m_{jj}$ closest in $\Delta R$ (`$m_{jj}$-in' and `$m_{jj}$-out') and the flavors of SS leptons (`$e^{\pm}e^{\pm}$', `$e^{\pm}\mu^{\pm}$', and `$\mu^{\pm}\mu^{\pm}$').
Events with $65\GeV < m_{jj} < 95\GeV$ are categorized in $m_{jj}$-in, and the others in $m_{jj}$-out. 
On the other hand, the $3\ell$ events are classified into three categories by the flavors of final-state leptons: zero, one, or two same-flavor opposite-sign (SFOS) lepton pairs.\footnote{
The $3\ell$ production processes can affect the SS category if one of the charged leptons is not detected, though the contribution is minor.
}

The ALP model has been analyzed by the CMS collaboration in the limited mass region, $200\GeV < m_{a} < 600\GeV$~\cite{CMS:2019mpq}.
We utilize their result for the validation of our analysis. 
There are two main differences between our approach and the experimental analysis.
Although the production cross section is evaluated at the next-to-leading order in Ref.~\cite{CMS:2019mpq}, our analysis is performed at the leading order. 
Furthermore, we employed the ${\rm CL}_{s}$ method as explained above.
We found that our upper bound on $g_{aWW}$ is weaker by $\sim40\%$ compared to those in Fig.~3 in Ref.~\cite{CMS:2019mpq}.\footnote{
We also evaluated the upper bounds without using the ${\rm CL}_{s}$ method by employing the mean value of Poisson distributions for the upper bound following Ref.~\cite{Craig:2018kne}.
The results become stronger by $\sim50\%$ than those in Ref.~\cite{CMS:2019mpq}.
In addition, higher-order effects can affect the results. 
Those effects as well as the analysis method could cause the discrepancy from the CMS result.
}
Therefore, the ${\rm CL}_{s}$ method is found to provide a conservative limit on $g_{aWW}$.

\subsection{ATLAS \texorpdfstring{$Z(\to \nu\bar{\nu})\gamma jj$}{} search} \label{sec:Znunugamma}

\begin{figure}[t]
 \centering
 \includegraphics[scale=0.25]{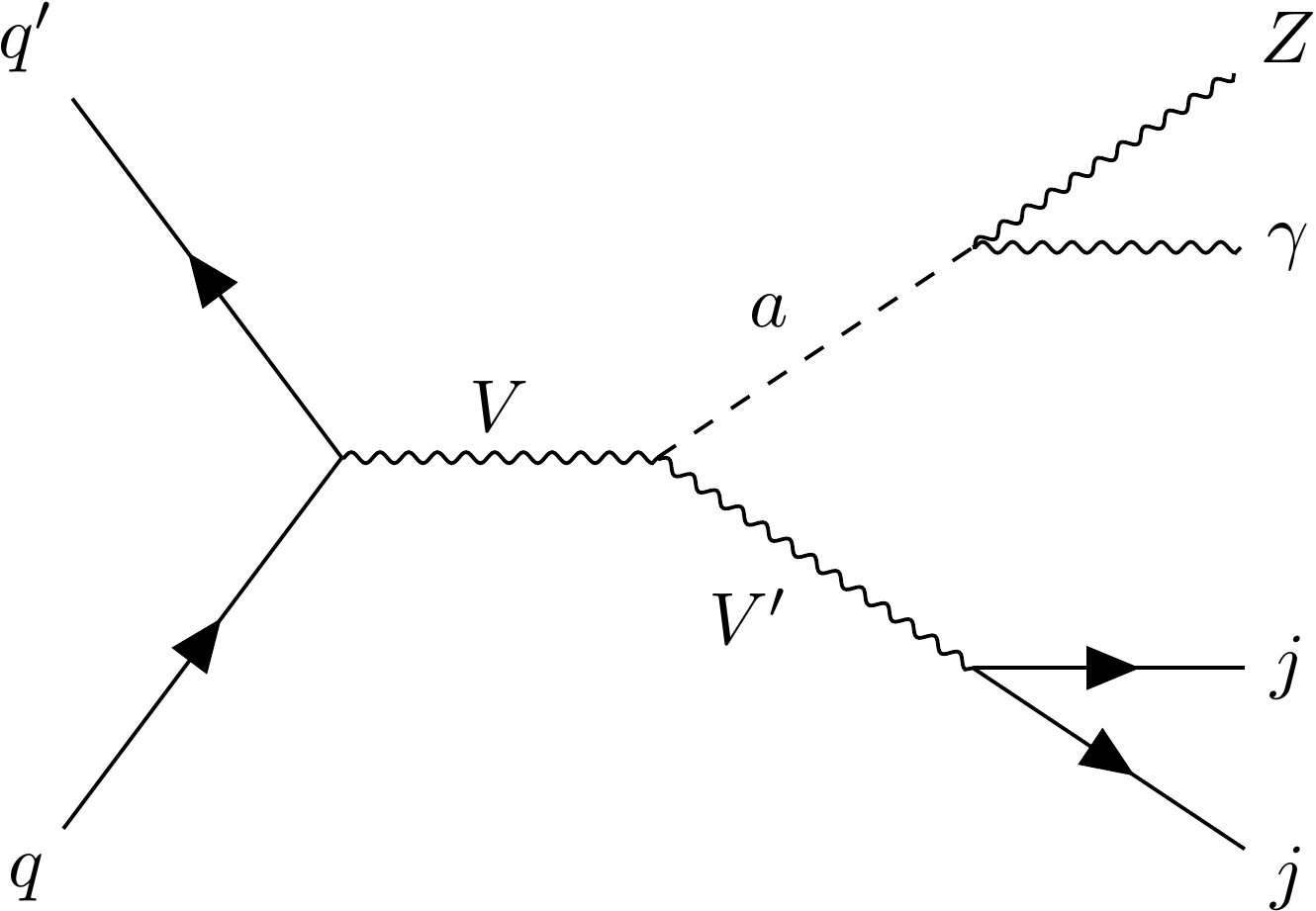}
 \hspace{8mm}
 \includegraphics[scale=0.25]{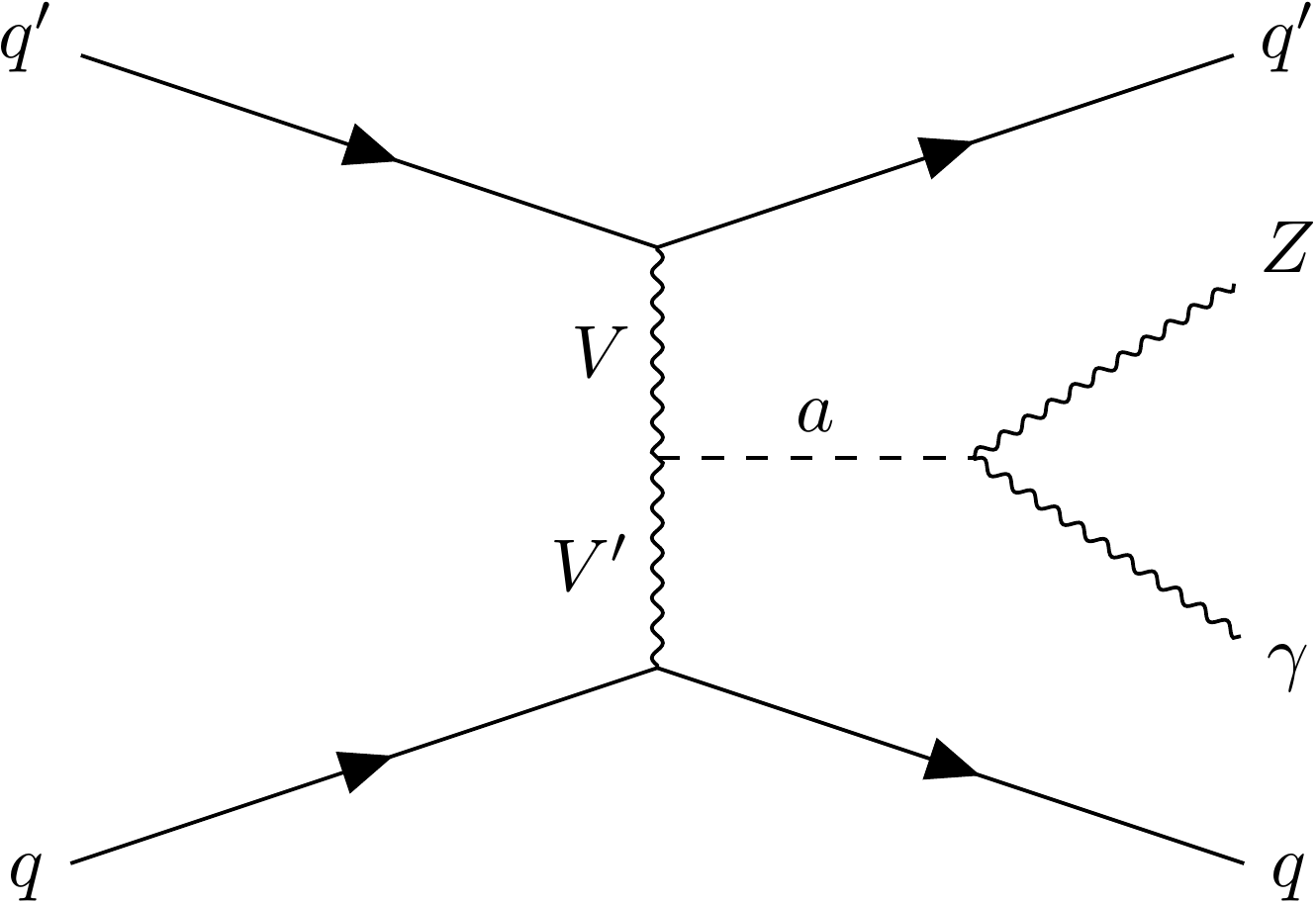} 
 \caption{Feynman diagrams for $pp\to Z\gamma jj$ production.}
 \label{fig: pp_zajj}
\end{figure}

The analysis is based on a data sample from the ATLAS collaboration at $\sqrt{s}=13\TeV$ with ${\cal L}_I = 139\,{\rm fb}^{-1}$~\cite{ATLAS:2021pdg}.
The signal process is $pp\to ajj \to Z(\to\nu\bar{\nu})\gamma jj$ (cf., Fig.~\ref{fig: pp_zajj}).
In the event production, off-shell contributions of $a$ and $Z$ are taken into account.
We reinterpret the experimental results using the signal categories aimed at vector-boson fusion (VBF) productions of a Higgs boson decaying into a photon $\gamma$ and massless invisible dark photon $\gamma_{D}$.

\begin{table}[t]
\scriptsize
\centering
\begin{tabular}{c c c c c c}
\hline
$\pt(j_{1})$ [$\GeV$] & $\pt(j_{2})$ [$\GeV$] & $\pt(j_{>2})$ [$\GeV$] & $N_{\rm jet}$ & $N_{\text{b-jet}}$ & $\abs{\Delta\eta_{jj}}$ \\
$>60$ & $>50$ & $>25$ & $2,3$ & $<2$ & $>3.0$ \\
\hline
$\eta(j_{1})\times\eta(j_{2})$ & $C_{3}$ & $|\Delta\phi(j_{i}, \vec{E}_{\rm T}^{\rm miss})|$ & $N_{\gamma}$ & $|\Delta\phi_{jj}|$ & $\pt(\gamma)$ [$\GeV$] \\
$<0$ & $<0.7$ & $>1.0$ & $1$ & $<2.0$ & $>15$, $<{\rm max}(110, 0.733\times m_{\rm T})$ \\
\hline
$m_{jj}$ [$\TeV$] & $E_{\rm T}^{\rm jets}$ [$\GeV$] & $E_{\rm T}^{\rm miss}$ [$\GeV$] & $C_{\gamma}$ & $N_{\ell}$ \\
$>0.25$ & $>130$ & $>150$ & $>0.4$ & $0$ \\
\hline
\end{tabular}
\caption{Kinematical cuts for the $Z(\to \nu\bar{\nu})\gamma jj$ events based on Ref.~\cite{ATLAS:2021pdg}.}
\label{tab:vvajj}
\end{table}

The event reconstruction and event selection criteria are based on Ref.~\cite{ATLAS:2021pdg}.
We modify the default \verb|DELPHES| card for ATLAS so that jet clustering is performed in the anti-$k_{T}$ algorithm with distance parameter $R=0.4$.
For other variables, we use the default values.
The kinematical cuts are summarized in Table~\ref{tab:vvajj}, where the photon centrality $C_{\gamma}$ is defined as
\begin{align}
    C_{\gamma} = \exp\left[-\frac{4}{[\eta(j_{1})-\eta(j_{2})]^{2}}\left(\eta(\gamma)-\frac{\eta(j_{1})+\eta(j_{2})}{2}\right)^{2}\right].
\end{align}
The centrality for a third leading jet $C_{3}$ can be obtained by replacing $\eta(\gamma)$ with $\eta(j_{3})$.
The transverse mass constructed from the photon and $E_{\rm T}^{\rm miss}$ is defined as
\begin{align}
    m_{\rm T}(\gamma, E_{\rm T}^{\rm miss}) = \sqrt{2\pt^{\gamma}E_{\rm T}^{\rm miss}\left[1-\cos(\phi_{\gamma}-\phi_{E_{\rm T}^{\rm miss}})\right]}\, .
\end{align}
The events are classified into ten `categories' by the invariant mass of the leading two jets $m_{jj}$ ($m_{jj}<1\TeV$ and $m_{jj}\geq1\TeV$) and $m_{\rm T}$ ($0\text{--}90$, $90\text{--}130$,  $130\text{--}200$, $200\text{--}350$, and $> 350\GeV$).

For the validation, we evaluated VBF productions of the Higgs boson decaying into $\gamma + \gamma_{D}$.
The signal events were produced at the next-to-leading order in Ref.~\cite{ATLAS:2021pdg} by using \texttt{POWHEG v2}~\cite{Nason:2004rx, Frixione:2007vw, Alioli:2010xd, Nason:2009ai}, where the diagrams exchanging a weak boson at the $s$-channel are discarded.
For comparison, we generated the events at the leading order using \verb|MadGraph5_aMC@NLO 3.5.2| while excluding the $s$-channel diagrams.\footnote{
We checked that the results are almost unchanged even if the $s$-channel diagrams are included. 
}
We found that our results are consistent with those of Auxiliary table~3 in HEPData\footnote{\href{https://www.hepdata.net/record/ins1915357}{https://www.hepdata.net/record/ins1915357}}, but larger by $20\text{--}40\%$ than those in Table~9 in Ref.~\cite{ATLAS:2021pdg}.

\subsection{ATLAS \texorpdfstring{$Z(\to \ell^+\ell^-)\gamma$}{} search} \label{sec:Zllgamma}

The analysis is based on a data sample from the ATLAS collaboration at $\sqrt{s}=13\TeV$ with ${\cal L}_I = 140\,{\rm fb}^{-1}$~\cite{ATLAS:2023wqy}.
In contrast to the previous processes, the experimental analysis is aimed at searching for new resonances in the $Z\gamma$ final state.
We recast the experimental results specified for CP-even spin-0 productions, where the upper bound on the production cross section times a branching ratio of the resonance decaying into $Z\gamma$ is provided. 
The result is given in the mass range from $220\GeV$ to $3400\GeV$.
The signal process is $pp\to ajj \to Z(\to\ell^{+}\ell^{-})\gamma jj$ (cf., Fig.~\ref{fig: pp_zajj}).
Since it is the resonance signature that is of interest, the ALPs are assumed to be produced on-shell, and off-shell contributions are discarded. 

\begin{table}[t]
\scriptsize
\centering
\begin{tabular}{c c c c c} 
\hline\hline 
& Muon & Electron & Electron as photon & Photon \\
\hline
$\pt$ & $>10\GeV$ & $>10\GeV$ & $>50\GeV$ & $>15\GeV$ \\
\hline
\multirow{2}{*}{$|\eta|$} & \multirow{2}{*}{$<2.7$} & $<2.47$ & $<2.47$ & $<2.37$ \\
& & Exclude [1.37, 1.52]  & Exclude [1.37, 1.52] & Exclude [1.37, 1.52] \\
\hline
$\Delta R(\text{track},\gamma)$ & & & $< 0.1$ & \\
\hline
$ee$ or $\mu\mu$ pair & \multicolumn{2}{c}{$\ge 2$, opposite charge} & & \\
\multirow{2}{*}{$e\gamma$ pair} & & & $\Delta R(e,\gamma) < 1 $ & \\
& & & $|\pt^{e} - \pt^{\gamma}|/\pt^{e\text{ or }\gamma} > 5\% $ & \\
\hline
Categorization &\multicolumn{4}{c}{lepton pair closest to $m_{Z}=91.2\GeV$, decide if electron channel or muon channel}\\
\hline
\multirow{3}{*}{Event selections}& \multicolumn{4}{c}{$|m_{\ell\ell}^{\text{corrected}}-m_{Z}|<15\GeV$, $m_{Z}=91.2\GeV$} \\
& \multicolumn{4}{c}{$\pt^{\gamma}/m_{Z\gamma}>0.2$;  signal region: $100 < m_{Z\gamma} < 4000\GeV$}\\
\hline
\hline
\end{tabular}
\caption{Kinematical cuts for the $Z(\to \ell^{+}\ell^{-})\gamma jj$ events, which are based on Ref.~\cite{ATLAS:2023wqy} but the cut for $m_{Z\gamma}$ (see the text).}
\label{tab:llajj}
\end{table}

The event reconstruction and selection criteria are based on Ref.~\cite{ATLAS:2023wqy}.
We use the default \verb|DELPHES| card for ATLAS.
The kinematical cuts are summarized in Table~\ref{tab:llajj}.
Here, we put a looser cut on the $Z\gamma$ invariant mass $m_{Z\gamma}$ ($100\GeV < m_{Z\gamma} < 4000\GeV$) than that in Ref.~\cite{ATLAS:2023wqy} ($200\GeV < m_{Z\gamma} < 3500\GeV$).
This is because we evaluated $A \times \epsilon$ for CP-even spin-0 resonances by ourselves and found that the result was inconsistent with the experimental one (cf., Fig.~1 in Ref.~\cite{ATLAS:2023wqy}).
Among the kinematical cuts, we noticed that the results are sensitive to the choice of $m_{Z\gamma}$.
It may be affected by reconstruction resolutions, and instead of refining the smearing function implemented in \verb|DELPHES|, we simply loosen the cut for $m_{Z\gamma}$ such that the experimental result is reproduced.\footnote{
The modification is applied just to reproduce the experimental result for the CP-even spin-0 resonances; it is not applied to the background events but to the signal processes only.
The number of signal events is insensitive to the choice of the $m_{Z\gamma}$ cut as long as it is wide enough. 
}
Then, the upper limit on the number of signal events is obtained as
\begin{align}
    n_{\rm up}^{\rm sig} = 
    \sigma(pp\to X\to Z\gamma)_{\rm up}
    \cdot {\cal L}_{I}
    \cdot \left.A\times\epsilon\right|_{\text{Spin-0}},
\end{align}
where $\sigma(pp\to X\to Z\gamma)_{\rm up}$ is the observed upper limit on the production cross section times branching ratio of the spin-0 resonance in Ref.~\cite{ATLAS:2023wqy}.
On the other hand, the number of signal events is obtained for the ALP as
\begin{align}
    n^{\rm sig} =
    \sigma(pp\to a j j\to Z\gamma jj)
    \cdot {\cal L}_{I}
    \cdot \left.A\times\epsilon\right|_{\text{ALP}}.
\end{align}
Since the cross section is from on-shell productions of ALPs, it is proportional to $g_{aWW}^{2}$, and we obtain an upper limit on $g_{aWW}$ from $n^{\rm sig} < n_{\rm up}^{\rm sig}$.

\section{Results} \label{sec: numerical results}

\begin{figure}[t]
 \centering
 \includegraphics[scale=0.6]{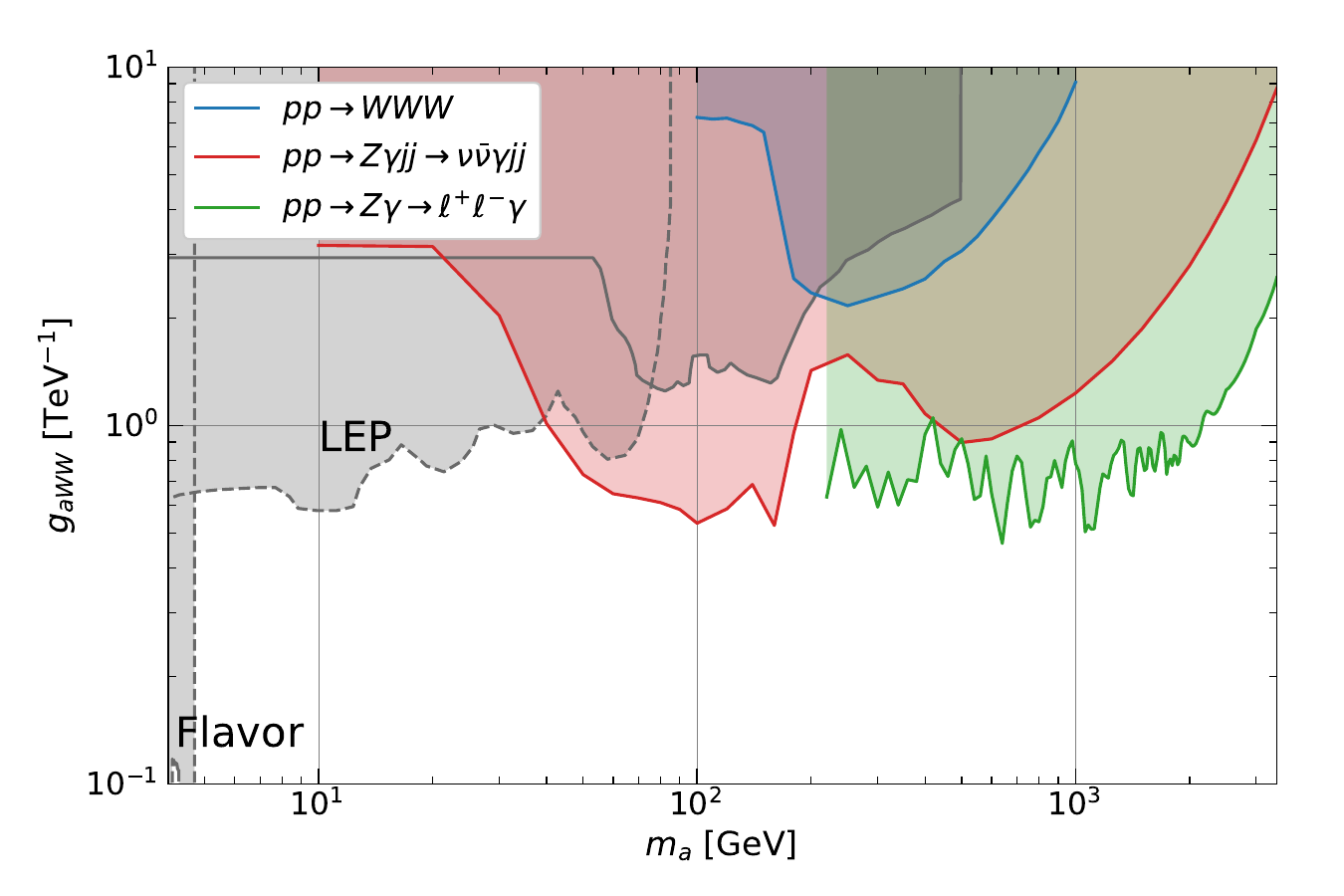} 
 \caption{Observed 95\% upper limits on $g_{aWW}$ as a function of ALP mass obtained by the multi-boson searches for $pp\to WWW$ (blue), $pp\to Z\gamma\to \nu\bar{\nu}\gamma$ (red), and $pp\to Z\gamma\to \ell^{+}\ell^{-}\gamma$ (green) at the LHC with $\sqrt{s}=13\TeV$.
 Here, the intermediate ALP, $Z$, and $W$ bosons can be off-shell (see text for details).
 The gray regions surrounded by dashed lines are excluded by LEP constraints and flavor limits.
 The gray region surrounded by a solid line indicates the previous LHC constraint~\cite{Craig:2018kne}.}
 \label{fig: gaWW_ma}
\end{figure}

In Fig.~\ref{fig: gaWW_ma}, we show the observed 95\% upper limit on $g_{aWW}$ as a function of ALP mass $m_{a}$.
Here, $\Lambda = 4\pi f_{a} = 4\pi\times10\TeV$ is chosen to evaluate the branching ratios of loop-induced ALP decays. 
The bounds are obtained by reinterpreting the multi-boson searches for $pp\to WWW$ (blue), $pp\to Z\gamma\to \nu\bar{\nu}\gamma$ (red), and $pp\to Z\gamma\to \ell^{+}\ell^{-}\gamma$ (green).
They are compared with the other constraints; the gray regions surrounded by dashed lines are excluded by LEP constraints and flavor limits.\footnote{
The process $e^+e^- \to \Upsilon \to \gamma a$ with $a \to \text{SM fermions}$ is studied in Ref.~\cite{DiLuzio:2024jip} based on the data at BaBar, Belle, and Belle II.
It is sensitive to the ALP contribution with $m_a \lesssim 10\GeV$.
We found that their constraints on the current model are much weaker than the LEP and LHC bounds in Fig.~\ref{fig: gaWW_ma}, because the ALP is coupled to the SM fermions via radiative corrections. 
}
We also show the previous LHC constraint~\cite{Craig:2018kne} by the gray region surrounded by a solid line.

It is found that the best limit is obtained by $pp\to Z\gamma\to \ell^{+}\ell^{-}\gamma$ (green) for $220\GeV < m_{a} < 3400\GeV$, for which the ATLAS result for new resonance searches~\cite{ATLAS:2023wqy} is used.
The scattering proceeds via on-shell ALP production associated with two jets, $pp\to ajj \to Z(\to\ell^{+}\ell^{-})\gamma jj$ (See Fig.~\ref{fig: pp_zajj}).
Here, the $Z$ boson as well as the ALP is produced on-shell.

For $m_{a} < 220\GeV$, the limit is given by $pp\to \nu\bar{\nu} \gamma jj$ (red).\footnote{
Here, we generate the ALP events including the $s$-channel diagrams (cf., the left panel of Fig.~\ref{fig: pp_zajj}). The results are insensitive to the contribution due to the kinematical cuts.
Also, the production cross section for $pp \to a (\to \nu\bar\nu) \gamma jj$ is suppressed by ${\rm BR}(a \to \nu\bar\nu) \simeq 0$. 
}
If the mass is larger than $\sim 100\GeV$, the signal events are dominated by $pp\to ajj \to Z(\to\nu\bar{\nu})\gamma jj$, where both the ALP and $Z$ boson are on-shell. 
On the other hand, in the smaller mass regions either $a$ or $Z$ would be off-shell.
We found that the ALP is favored to be on-shell for $m_{a} \gtrsim 30\GeV$, \ie, the process is $pp\to ajj,\, a \to Z^*\gamma \to \nu\bar{\nu}\gamma$.
Nonetheless, there are non-negligible contributions from the off-shell ALP productions, which are taken into account in our analysis.
It is noted that the loop-induced ALP decays, especially $a \to b\bar b$, are taken into account in evaluating the signal events because they dominate the ALP total decay width for $m_{a}\lesssim m_{Z}$ (see Fig.~\ref{fig: ALP_decay_gaAA0gaWW01gaGG0}). 
It is found that the LHC results supersede the LEP constraint for $m_{a} > 40\GeV$.
For lighter ALPs, the upper limit on $g_{aWW}$ is degraded because the branching ratio of $a \to Z^* \gamma \to \nu\bar\nu\gamma$ is suppressed.
For $m_{a} \lesssim 30\GeV$ the signal events are dominated by the off-shell ALP productions, and the constraint on $g_{aWW}$ eventually becomes insensitive to the ALP mass.
In the plot, the result for $pp\to \nu\bar{\nu} \gamma jj$ is shown only in the region $m_{a} > 10\GeV$ because the on-shell ALP productions are of interest in this paper. 

Although the ALP decay is dominated by $a \to WW$ for $m_{a} \gtrsim 2 m_{W}$, we find that the measurement of $pp\to WWW$ (blue) provides a weaker limit on the model.
The signal events are dominated by on-shell productions of ALPs followed by decay into a pair of on-shell $W$ bosons ($pp\to a W,\, a \to WW$) for $m_{a} \gtrsim 2 m_{W}$.
The off-shell ALP contributions are relevant for lighter ALPs. 
The constraint on $g_{aWW}$ eventually becomes insensitive to the ALP mass, and the result is shown only for $m_{a} > 100\GeV$.

As mentioned above, the ALP is favored to be on-shell even below the $Z$-boson mass scale for $pp\to \nu\bar{\nu} \gamma jj$. 
In contrast, for $pp\to WWW$ the on-shell ALP production $pp\to aW,\, a\to WW^{*}$ does not dominate the signal events for $m_{a} \lesssim 2 m_{W}$.
This is probably because the branching ratio of $a\to WW^{*}$ is suppressed due to $a\to Z\gamma$ proceeding at tree level in the latter case.
In the former case ($pp\to \nu\bar{\nu} \gamma jj$), since the ALP total width is dominated by the loop-induced decay $a \to b\bar b$, the three-body decay $a \to Z^*\gamma \to \nu\bar{\nu}\gamma$ can contribute sizably even for $m_{a} \lesssim m_{Z}$.

Let us compare our results with previous works. 
In Ref.~\cite{Craig:2018kne}, the collider constraints on on-shell ALP productions have been studied with the LEP and LHC results. 
In particular, the $Z$-boson decays were investigated for the former, and the latter was based on the data at $\sqrt{s}=8\TeV$ and $\int\!\mathcal{L}\,dt\simeq 20\,{\rm fb}^{-1}$.
In comparison, we find that our result provides the best limit for $m_{a} > 40\GeV$. 
Besides, even $pp\to WWW$ gives a stronger bound than what is given in the literature for $m_{a} > 200\GeV$. 
In Ref.~\cite{Bonilla:2022pxu}, off-shell ALP production has been studied for LHC data at $\sqrt{s}=13\TeV$.
We found that our results are more severe than what is given in the literature for $m_{a} > 20\GeV$, where the off-shell productions are dominant, and the constraint for $m_{a} < 20\GeV$ is comparable with the previous one.

\section{Conclusion} \label{sec: conclusion}

In this paper, we have studied the photophobic ALP model in high-mass regions using the LHC Run-II results.
Since the ALP-photon coupling is suppressed, the model can avoid tight constraints from searches for di-photon final states.
The ALPs are predominantly coupled with electroweak gauge bosons, and we investigated the multi-boson final-state signatures at the LHC. 
We reinterpreted the LHC analyses of $pp\to WWW$, $pp\to Z\gamma\to \nu\bar{\nu}\gamma$, and $pp\to Z\gamma\to \ell^+\ell^-\gamma$. 
We found that the on-shell ALP productions via $pp \to Z\gamma \to \nu\bar{\nu}\gamma$ shows the best sensitivity in the mass range, $40 < m_{a} < 220\GeV$.
For the mass region $220 < m_{a} < 3400\GeV$, the ALP resonance search via $pp\to Z\gamma\to \ell^{+}\ell^{-}\gamma$ provides the most severe constrains.

\section*{Acknowledgements}
This work is supported by the Japan Society for the Promotion of Science (JSPS) Grant-in-Aid for Scientific Research B (No. 21H01086 [M.E.~and K.F.]), and JSPS KAKENHI Grants No.~22KJ3126 [M.A.] and No.~22K21350 [K.F.].

\appendix
\section{Case with ALP-gluon coupling} \label{sec: ALP-gluon coupling}

\begin{figure}[t]
 \centering
 \includegraphics[scale=0.6]{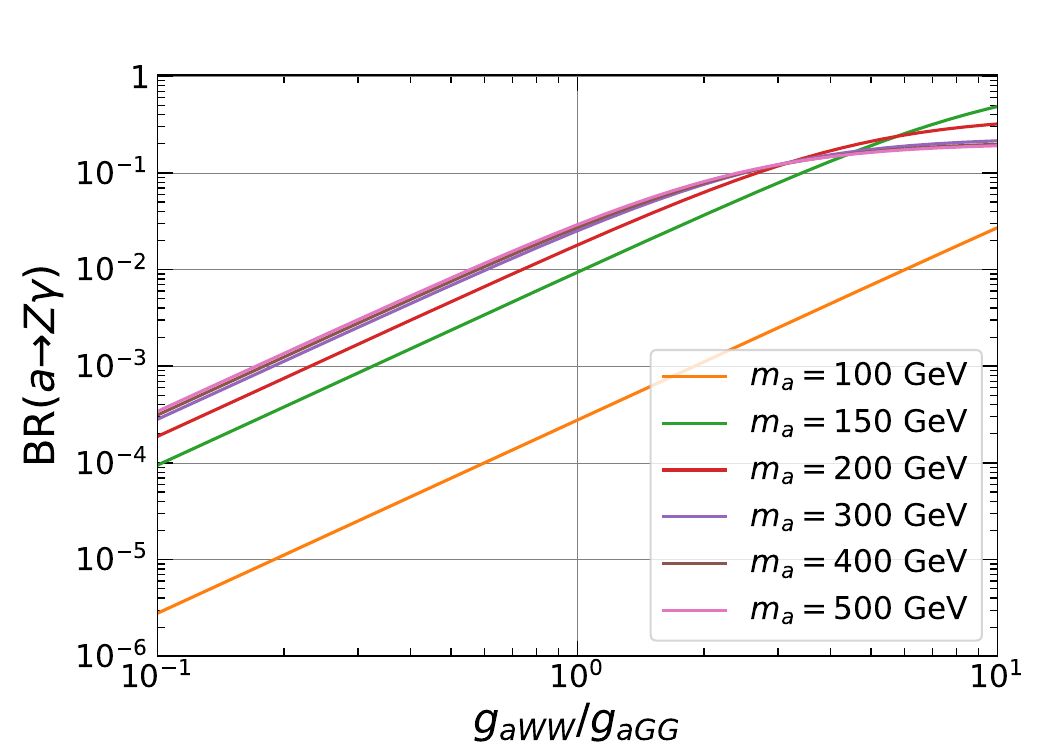}
 \caption{Branching ratios of the ALP decaying into $Z\gamma$ as a function of $g_{aWW}/g_{aGG}$ for $m_{a} = 100,\,150,\,200,\,300,\,400,\,500\GeV$ and $\Lambda = 4\pi f_{a} = 4\pi\times10\TeV$.}
 \label{fig: BRaToAZ_RgaWWgaGG_ma500}
\end{figure}

In this section, we discuss LHC signatures for the photophobic ALP in the presence of an ALP-gluon coupling at tree level.
The interaction is expressed as
\begin{align}
    {\cal L}_{\rm int} = -\frac{1}{4}g_{aGG} \, a G_{\mu\nu}^{a}\widetilde{G}^{a\mu\nu}.
\end{align}
The ALP can be produced from and decay back into a pair of gluons.
For the search via $pp\to Z\gamma\to \ell^{+}\ell^{-}\gamma$, we find that the production cross section of ALPs is dominated by the gluon fusion if $g_{aGG}$ is larger than $\sim 0.01 \times g_{aWW}$.
In the following, we assume that the gluon fusion process is a dominant ALP production channel.
Furthermore, the branching ratio of the ALP decaying into electroweak gauge bosons is affected by the ALP-gluon coupling.
The effective coupling with quarks in Eq.~\eqref{eq: cff_eff} is replaced as
\begin{align}
    g_{aff}^{\rm eff} &=
    3C_{F}\frac{\alpha_{s}}{4\pi}g_{aGG} \ln{\frac{\Lambda^{2}}{m_{f}^{2}}}
    +\frac{3}{4s_{W}^{2}}\frac{\alpha}{4\pi}g_{aWW} \ln{\frac{\Lambda^{2}}{m_{W}^{2}}}
    \notag \\ &\quad
    +\frac{3}{s_{W}c_{W}}\frac{\alpha}{4\pi}g_{aZ\gamma}Q_{f}
        \qty(I_{3}^{f}-2Q_{f}s_{W}^{2})
        \ln{\frac{\Lambda^{2}}{m_{Z}^{2}}}
    \notag \\ &\quad
    +\frac{3}{s_{W}^{2}c_{W}^{2}}\frac{\alpha}{4\pi}g_{aZZ}
        \qty(Q_{f}^{2}s_{W}^{4}-I_{3}^{f}Q_{f}s_{W}^{2}+\frac{1}{8})
        \ln{\frac{\Lambda^{2}}{m_{Z}^{2}}},
    \label{eq: cff_eff}
\end{align}
where $C_{F}=4/3$. Here, we keep the terms enhanced by $\ln{(\Lambda^2/m_f^2)}$ or $\ln{(\Lambda^2/m_{V}^2)}$ $(V=Z, W)$.
Subleading terms are given in Refs.~\cite{Bauer:2017ris, Bauer:2020jbp, Bonilla:2021ufe}, though they are irrelevant here.
For $m_{a} \gtrsim 3\GeV$, the hadronic decay rates are calculated perturbatively~\cite{Bauer:2017ris, Bauer:2020jbp, Bonilla:2021ufe, Arias-Aragon:2022iwl}.\footnote{
Meson productions should be considered for lighter cases~\cite{Bauer:2017ris, Aloni:2018vki}.
}
As for the hadronic decay rate, it is sufficient to evaluate Eq.~\eqref{eq: aToHadron} with the tree-level ALP-gluon coupling, \ie, $g_{aGG}^{\rm eff}=g_{aGG}$.
In Fig.~\ref{fig: BRaToAZ_RgaWWgaGG_ma500}, we show the branching ratio of $a\to Z\gamma$ as a function of $g_{aWW}/g_{aGG}$.
We find that the ratio decreases due to the hadronic contributions for $g_{aWW} \lesssim g_{aGG}$.
For $m_{a}\lesssim 2m_{W}$, the ratio increases for larger $g_{aWW}/g_{aGG}$.
On the other hand, the ratio is close to 0.2 at $g_{aWW}/g_{aGG} \sim 10$ for $m_{a}> 2m_{W}$, where the total width is dominated by $a \to WW$, and $a\to Z\gamma$ provides the next-to-largest contribution (see Fig.~\ref{fig: ALP_decay_gaAA0gaWW01gaGG0}).

\begin{figure}[t]
 \centering
 \includegraphics[scale=0.475]{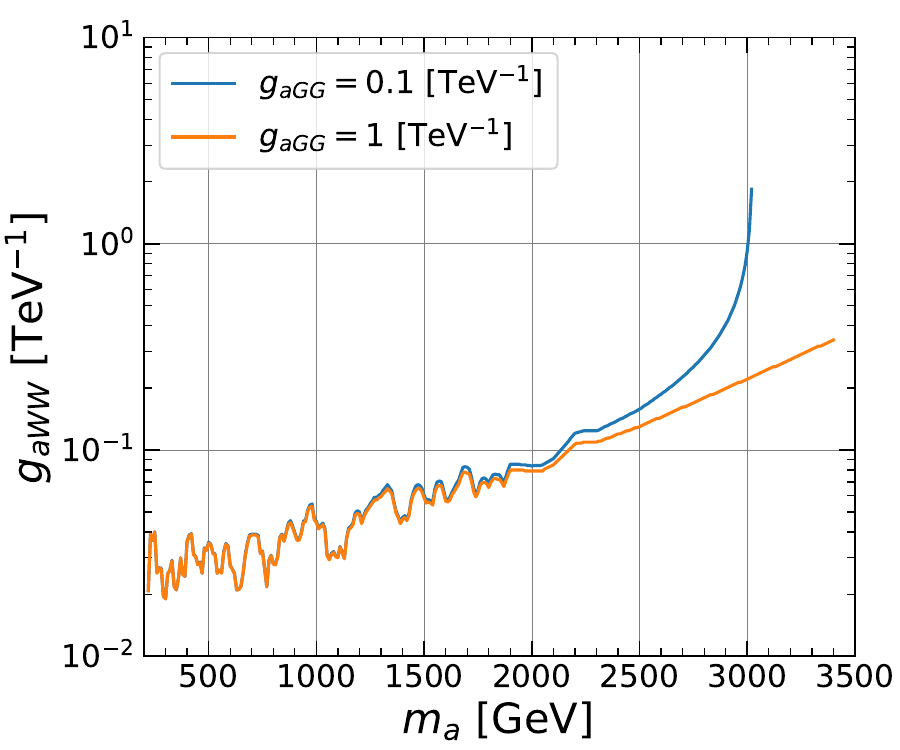}
 \includegraphics[scale=0.475]{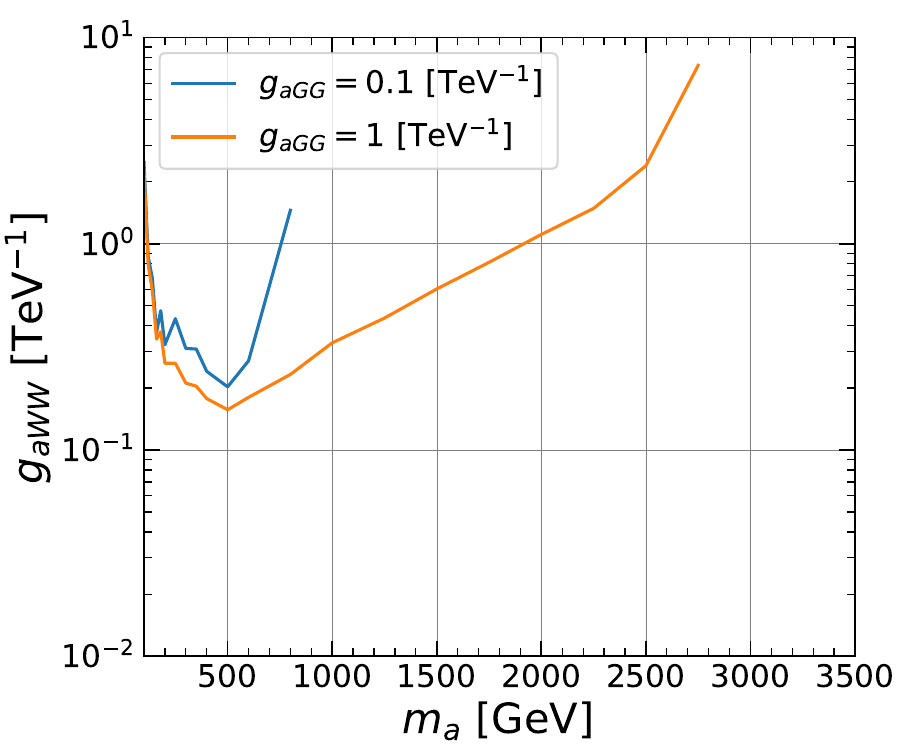}
 \caption{Observed 95\% upper limits on $g_{aWW}$ for $g_{aGG} = 0.1\TeV^{-1}$ (blue) and $1\TeV^{-1}$ (orange) as a function of $m_{a}$ obtained by the multi-boson searches for $pp\to Z\gamma\to \ell^{+}\ell^{-}\gamma$ (left) and $pp\to Z\gamma\to \nu\bar{\nu}\gamma$ (right).} 
 \label{fig: ObsLim_gaWW_w_gaGG}
\end{figure}

In Fig.~\ref{fig: ObsLim_gaWW_w_gaGG}, we show the observed 95\% upper limit on $g_{aWW}$ as a function of ALP mass $m_{a}$ for $g_{aGG} = 0.1\TeV^{-1}$ (blue) and $1\TeV^{-1}$ (orange).
The bounds are obtained by reinterpreting the experimental results for $pp\to Z\gamma\to \ell^{+}\ell^{-}\gamma$ (left) and $pp\to Z\gamma\to \nu\bar{\nu}\gamma$ (right).
In the left panel, the constraints are provided for $m_{a} > 220\GeV$. 
We find that the results are insensitive to $g_{aGG}$ for $m_{a} \lesssim 2\TeV$.
This is because the cross section is given by the on-shell ALP production, $gg \to a \to Z\gamma,\, Z \to \ell^{+}\ell^{-}$, and is proportional to $\sigma(gg \to a) \times {\rm Br}(a \to Z\gamma) \times {\rm Br}(Z \to \ell^{+}\ell^{-})$.
Note that $\sigma(gg \to a)$ is proportional to $g_{aGG}^2$.
If $g_{aWW}$ is small enough, since ${\rm Br}(a \to Z\gamma)$ is scaled by $(g_{aWW}/g_{aGG})^2$ (see Fig.~\ref{fig: BRaToAZ_RgaWWgaGG_ma500}), the cross section becomes insensitive to $g_{aGG}$. 
For larger $m_{a}$ the constraints lie above $g_{aWW} \gtrsim 0.1\TeV^{-1}$, and ${\rm Br}(a \to Z\gamma)$ is not scaled by $(g_{aWW}/g_{aGG})^2$ for $g_{aGG} = 0.1\TeV^{-1}$.
The constraint line terminates at $m_{a} \simeq 3\TeV$ for $g_{aGG} = 0.1\TeV^{-1}$.
As the ALP mass is increased, the production cross section decreases, and a larger branching ratio is required to saturate the upper bound on the number of signal events. 
However, since ${\rm Br}(a \to Z\gamma)$ cannot be too large, LHC searches do not have a sensitivity above the ALP mass scale. 

In the right panel, the $pp\to Z\gamma\to \nu\bar{\nu}\gamma$ constraints are shown for $m_{a} > 100\GeV$.
Here, the on-shell ALP production $gg \to a \to Z\gamma,\, Z \to \nu\bar{\nu}$ dominates the signal events.
Since the constraint is weaker than $g_{aWW} = 0.1\TeV^{-1}$, the branching ratio of $a \to Z\gamma$ is not scaled by $(g_{aWW}/g_{aGG})^2$ for $g_{aGG} = 0.1\TeV^{-1}$.
For heavier ALPs, since the production cross section decreases, the constraint line terminates at $m_{a} \simeq 800\GeV$ $(2700\GeV)$ for $g_{aGG} = 0.1\TeV^{-1}$ ($1\TeV^{-1}$).

In summary, although $g_{aGG} = 0.1\TeV^{-1}$ and $1\TeV^{-1}$ are likely to be too large to be generated by renormalization group evolution because the contribution arises at the two-loop level in the presence of $g_{aWW}$ (see the discussion in Sec.~\ref{sec: ALP_model}), the ALP-gluon coupling can affect the LHC signatures significantly if it is introduced. 
It is found that if $g_{aGG}$ is sizable the LHC measurements can provide better sensitivities than the original photophobic ALP model.

\bibliographystyle{utphys28mod}
\bibliography{references}
\end{document}